\documentclass[12pt]{iopart}
\usepackage{iopams}
\usepackage{amssymb}
\usepackage{graphicx}

\begin{document}

\title{Collapse instability of solitons in the nonpolynomial Schr\"{o}dinger
equation with dipole-dipole interactions}
\author{G. Gligori\'c$^1$, A. Maluckov$^2$, Lj. Had\v zievski$^1$, and B. A.
Malomed$^3$}

\begin{abstract}
A model of the Bose-Einstein condensate (BEC) of dipolar atoms, confined in
a combination of a cigar-shaped trap and optical lattice acting in the axial
direction, is studied in the framework of the one-dimensional (1D)
nonpolynomial Schr\"{o}dinger equation (NPSE) with additional terms
describing long-range dipole-dipole (DD) interactions. The NPSE makes it
possible to describe the collapse of localized modes, which was
experimentally observed in the self-attractive BEC confined in tight traps,
in the framework of the 1D description. We study the influence of the DD
interactions on the dynamics of bright solitons, especially as concerns
their collapse-induced instability. Both attractive and repulsive contact
and DD interactions are considered. The results are summarized in the form
of stability/collapse diagrams in a respective parametric space. In
particular, it is shown that the attractive DD interactions may prevent the
collapse instability in the condensate with attractive contact interactions.
\end{abstract}

\maketitle

\address{$^1$ Vin\v ca Institute of Nuclear Sciences, P.O. Box 522,11001 Belgrade,
Serbia \\
$^2$ Faculty of Sciences and Mathematics, University of Ni\v s, P.O. Box
224, 18001 Ni\v s, Serbia \\
$^3$ Department of Physical Electronics, School of Electrical Engineering,
Faculty of Engineering, Tel Aviv University, Tel Aviv 69978, Israel} %
\ead{goran79@vinca.rs}


\section{Introduction}

In the mean-field approximation, the dynamics of Bose-Einstein
condensates (BECs) obeys the 3D Gross-Pitaevskii equation (GPE)
\cite{BEC}, from which an effective 1D equation can be derived, in
various settings, for condensates trapped in prolate traps
\cite{PerezGarcia98}-\cite{Napoli2}. In the simplest case, which
corresponds to a sufficiently low BEC\ density, the reduction of the
3D equation for the BEC trapped in the ``cigar-shaped" configuration
leads to the one-dimensional (1D) cubic nonlinear Schr\"{o}dinger
equation (NLSE) \cite{Panos}. The main restriction on the use of
this equation is its inability to describe the onset of collapse of
localized states, which was theoretically predicted in the 3D
setting and experimentally observed in the self-attractive BEC \cite%
{Strecker02}, \cite{Cornish}. However, without imposing the
constraint of a very low density, the reduction of the 3D GPE leads
to the 1D equation with a \textit{nonpolynomial} nonlinearity, alias
the nonpolynomial Schr\"{o}dinger equation (NPSE)
\cite{sala1,Canary}. The 1D NPSE with the attractive sign of the
nonlinearity enables the description of the collapse dynamics and
produces results which are corroborated by direct simulations of the
underlying 3D GPE \cite{various}.

On the other hand, it is known that the mean-field description of BECs
trapped in a very deep optical-lattice (OL) potential can be well described
by the corresponding discrete equations. In particular, discrete forms of
the 1D GPE with the cubic nonlinearity \cite{DNLS-BEC,gpe,DNLS-BEC-review},
and of the 1D NPSE \cite{luca} have been studied in detail. Basic features
of the 1D continual equations describing BECs trapped in deep OLs find their
counterparts in the discrete models. In particular, the ability of the
continual 1D NPSE to capture the onset of the collapse is also shared by the
corresponding discrete equations.

A new variety of the BEC dynamics, which is dominated by long-range
(nonlocal) interactions, occurs in dipolar condensates, which may be formed
by magnetically polarized $^{52}$Cr atoms \cite{Cr}, dipolar molecules \cite%
{hetmol}, or atoms with electric dipole moments induced by an external field
\cite{dc}. A review of dynamical effects produced by the dipole-dipole (DD)
interactions in condensates can be found in Ref. \cite{review}. In
particular, conditions for the stability of the trapped dipolar BEC against
collapse were studied in detail \cite{collapse}. A possibility of the
creation of 2D solitons in dipolar condensates was predicted too. Namely,
isotropic solitons \cite{Pedri05} and solitary vortices \cite{Ami2} may
exist if the sign of the dipole-dipole (DD) interaction is inverted by means
of rapid rotation of the dipoles \cite{reversal}. On the other hand, stable
anisotropic solitons can be supported by the ordinary DD interaction, if the
dipoles are polarized in the 2D plane \cite{Ami1}. Solitons supported by
nonlocal interactions were also predicted and realized in optics, making use
of the thermal nonlinearity \cite{Krolik}.

A natural extension of the consideration of the dipolar BEC includes the OL
potential, which, in the discrete limit, leads to the model with the
long-range DD interactions between lattice sites \cite{nash,nash-novi,Santos}%
. Very recently, 1D solitons have been studied in the framework of the
continual cubic GPE, assuming the competition of local and nonlocal DD
interactions, with or without the OL potential \cite{competing}. In the
discrete limit, 1D solitons supported by the DD interactions were recently
studied too, in models with both the cubic \cite{nash} and nonpolynomial
\cite{nash-novi} onsite nonlinearity. In particular, the latter work
predicts a possibility of suppressing the collapse by means of the
long-range DD forces. The main purpose of the present work is to investigate
the influence of the DD interactions on the collapse dynamics in the 1D
continual NPSE, which is the most adequate setting for the study of the
onset and suppression of the collapse in the dipolar BEC loaded into a
cigar-shaped trap.

The paper is structured as follows. The model equation which includes the
nonpolynomial local nonlinearity and nonlocal DD interactions, is formulated
in Section II, where we also outline numerical techniques that we use in
this work. Focusing on the study of fundamental bright solitons, we report
basic results for their existence and stability in Section III. The core
part of paper is Section IV, where we study the influence of the DD
interaction on the solitons' collapse. The paper is concluded by Section V.

\section{The model}

The dynamics of a BEC at zero temperature is accurately described by the 3D
Gross-Pitaevskii equation (3D GPE) \cite{BEC}. When the condensate is
confined by a harmonic potential with frequency $\omega _{\bot }$ and
respective length $a_{\bot }=\left( \hbar /m\omega _{\bot }\right) ^{1/2}$
in the transverse plane, and by generic potential $V\left( z\right) $ in the
axial direction, it was shown in Ref. \cite{sala1} that the corresponding 3D
GPE can be reduced to the 1D NPSE for wave function $\psi \left( z,t\right) $%
, which is subject to normalization $\int_{-\infty }^{+\infty }\left\vert
\psi \left( z\right) \right\vert ^{2}dz=1$. The equation includes the OL
potential, with depth $V_{0}$ and wavenumber $K$, and the long-range DD (cf.
the 1D equations introduced in Refs. \cite{nash,competing}):%
\begin{eqnarray}
i\frac{\partial \psi }{\partial t} &=&-\frac{1}{2}\frac{\partial ^{2}\psi }{%
\partial z^{2}}+V_{0}\sin ^{2}\left( Kz\right) \psi +\frac{1-\frac{3}{2}%
\gamma \left\vert \psi \right\vert ^{2}}{\sqrt{1-\gamma \left\vert \psi
\right\vert ^{2}}}\psi  \nonumber \\
&&+G\psi (z)\int_{-\infty }^{+\infty }\frac{|\psi (z^{\prime })|^{2}}{%
|z-z^{\prime }|^{3}}dz^{\prime }.  \label{NPSE}
\end{eqnarray}%
Here, $\gamma =-2Na_{s}\sqrt{m\omega _{\bot }/\hbar }$ is the effective
strength of the local interaction, with $N$ the total number of atoms in the
condensate, and $a_{s}$ the scattering length of atomic collisions ($a_{s}<0$
corresponds to attraction) \cite{sala1}. Further, $G=g\left( 1-3\cos
^{2}\theta \right) $ is the coefficient which defines the DD interaction,
where $g$ is a positive coefficient, and $\theta $ the angle between the $z$
axis and the orientation of the dipoles.

Replacing wave function $\psi $ by $f\equiv \sqrt{\left\vert \gamma
\right\vert }\psi $, we transform Eq.(\ref{NPSE}) into a normalized form,%
\begin{eqnarray}
i\frac{\partial f}{\partial t} &=&-\frac{1}{2}\frac{\partial ^{2}f}{\partial
z^{2}}+V_{0}\sin ^{2}\left( Kz\right) f+\frac{1-\frac{3}{2}\aleph \left\vert
f\right\vert ^{2}}{\sqrt{1-\aleph \left\vert f\right\vert ^{2}}}f  \nonumber
\\
&&+\Gamma f(z)\int_{-\infty }^{+\infty }\frac{|f(z^{\prime })|^{2}}{%
|z-z^{\prime }|^{3}}dz^{\prime },  \label{NPSE1}
\end{eqnarray}%
where $\aleph =\mathrm{sgn}(\gamma )$ is the sign of the local interaction ($%
\aleph =+1$ corresponds to the attraction), and $\Gamma \equiv G/\left\vert
\gamma \right\vert $ measures the relative strength of the DD and contact
interactions.

One case of obvious interest for the 1D setting is that when the dipoles are
aligned with the $z$ axis, i.e., $\theta =0$ and $\Gamma =-2g/\left\vert
\gamma \right\vert <0$, which means that the DD interaction is attractive.
Another relevant case corresponds to the dipoles oriented perpendicular to
the $z$ axis, i.e. $\theta =\pi /2$ and $\Gamma =g/\left\vert \gamma
\right\vert >0$, which implies the repulsive DD interaction. Thus, the sign
of $\Gamma $ in the present model defines the character of the DD
interaction.

Stationary solutions to Eq. (\ref{NPSE1}), with chemical potential $\mu $,
are sought for as $f=u\exp (-i\mu t)$, with function $u$ satisfying the
stationary equation,
\begin{eqnarray}
\mu u &=&-\frac{1}{2}\frac{\partial ^{2}u}{\partial z^{2}}+V_{0}\sin
^{2}\left( Kz\right) u+\frac{1-\frac{3}{2}\aleph \left\vert u\right\vert ^{2}%
}{\sqrt{1-\aleph \left\vert u\right\vert ^{2}}}u  \nonumber \\
&&+\Gamma u(z)\int_{-\infty }^{+\infty }\frac{\left\vert u\left( z^{\prime
}\right) \right\vert ^{2}}{\left\vert z-z^{\prime }\right\vert ^{3}}%
dz^{\prime }.  \label{stationary}
\end{eqnarray}%
To present the results for soliton families, we will use the norm defined by
\begin{equation}
P=\int_{-\infty }^{+\infty }\left\vert u\left( z\right) \right\vert ^{2}dz
\label{Norm}
\end{equation}%
(according to the above definitions, $P$ is identical to $\left\vert \gamma
\right\vert ,$ but parameter $\gamma $ is not used below).

Experimentally adjustable coefficients in this model are the relative
strength of the DD/contact interactions, $\Gamma $, and the norm of the
stationary wave function, $P\sim N\left\vert a_{s}\right\vert \sqrt{m\omega
_{\bot }/\hbar }$ \cite{gpe,luca,nash}. In particular, $P\sim 1$ corresponds
to $\sim 1000$ atoms in the $^{52}$Cr condensate \cite{gpe,luca}. Actually, $%
\Gamma $ can be made both positive and negative, and its absolute value may
be altered within broad limits by means of the Feshbach resonance (without
the application of the Feshbach resonance, $\left\vert \Gamma \right\vert
\simeq 0.15$ in the condensate of chromium atoms) \cite{Cr,experim1}.

Stationary equation (\ref{stationary}) was solved numerically by means of an
algorithm based on the shooting method. We restrict the analysis to
fundamental solitons with a single maximum. Therefore, we chose the
following boundary conditions in the shooting procedure: the first
derivative of the wave function must be zero at the point of the maximum
amplitude, and the wave function must decay exponentially at infinity. Note
that the integral term in Eq. (\ref{stationary}) has the form of a
convolution \cite{goral},
\begin{equation}
\int_{-\infty }^{+\infty }V(z-z^{\prime })\left\vert u\left( z^{\prime
}\right) \right\vert ^{2}dz^{\prime }\equiv V(z)\times \left\vert u\left(
z\right) \right\vert ^{2},  \label{convolution}
\end{equation}%
hence its Fourier transform can be calculated as a product of the Fourier
images of the interaction potential and local density, $\left\vert u\left(
z\right) \right\vert ^{2}$. To calculate the integral term more accurately,
we adopted an iterative shooting procedure. For the first iteration, we
assumed that the potential of the interaction in the Fourier space is
approximated by a constant \cite{goral}. In each next iteration, we used the
previous solution to calculate the convolution integral numerically (\ref%
{convolution}).

Time-dependent equation (\ref{NPSE1}) was solved by means of the split-step
Fourier method \cite{splitfft}. To that end, Eq. (\ref{NPSE1}) was cast into
an operator form,
\begin{equation}
\frac{\partial f}{\partial t}=i(\hat{V}+\hat{D}+\hat{L}+\hat{N})f\,,
\label{split}
\end{equation}%
where $\hat{V}\equiv -V_{0}\sin ^{2}\left( Kz\right) $, $\hat{D}\equiv
-\Gamma \int_{-\infty }^{+\infty }|f(z^{\prime })|^{2}/|z-z^{\prime
}|^{3}dz^{\prime }$, $\hat{L}\equiv \partial ^{2}/(2\partial z^{2})$, and $%
\hat{N}\equiv -\left( 1-\aleph \left\vert f\right\vert ^{2}\right) ^{-1/2}%
\left[ 1-(3/2)\aleph \left\vert f\right\vert ^{2}\right] $. The split-step
Fourier method was implemented through independent actions of all of the
operators, $\hat{V}$, $\hat{D}$, $\hat{L}$ and $\hat{N}$, within time step $%
\tau $ of the integration scheme, $f(t+\tau ,z)=e^{-i\tau \hat{D}}e^{-i\tau
\hat{V}}e^{-i\tau \hat{L}}e^{-i\tau \hat{N}}f(t,z)$ (we fixed $\tau =10^{-4}$%
).

\section{Fundamental solitons}

\begin{figure}[tbp]
\center\includegraphics [width=7.7cm]{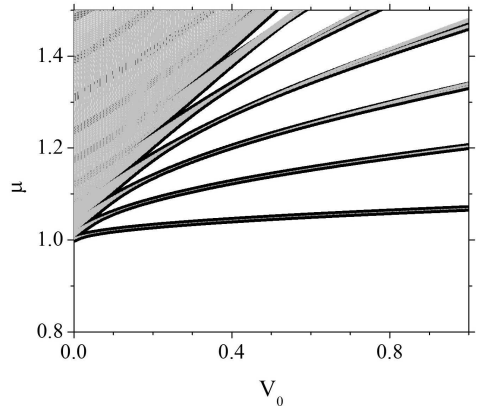}
\caption{The bandgap diagram for the linearized model. Shaded regions depict
the Bloch bands, separated by the gaps, where gap solitons can exist in the
nonlinear system. Solid lines depict band edges, which correspond to
periodic Bloch waves. The semi-infinite gap is located below the lowest
band. }
\label{fig1}
\end{figure}

Stationary wave function $u(z)$ was found from the solution of stationary
equation (\ref{stationary}). In the case of the noninteracting condensate ($%
\gamma =0$), and without the DD interactions ($\Gamma =0$), equation (\ref%
{stationary}) is tantamount to the Mathieu equation,
\begin{equation}
\frac{d^{2}u}{dy^{2}}-\left[ 2q\cos \left( 2y\right) -p\right] u=0,
\label{kivshar1}
\end{equation}%
where $y\equiv Kz$, $q\equiv -V_{0}/2K^{2}$, and $p\equiv 2\left(
q-1/K^{2}+\mu /K^{2}\right) $, hence the well-known bandgap diagram for the
Mathieu equation (see, e.g., Ref. \cite{kivshar-p}) can be used. That
diagram, translated into the notation adopted above, is displayed in Fig. %
\ref{fig1}. The nonlinear localization of matter waves in the form of gap
solitons may occur in gaps of the linear spectrum. In the case of the local
attractive nonlinearity, fundamental solitons populate the semi-infinite gap.

\begin{figure}[tbp]
\center\includegraphics [width=7.7cm]{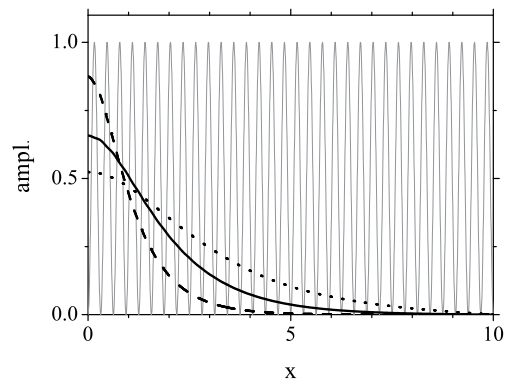}
\caption{Fundamental solitons for three different cases: in the absence of
the DD interaction ($\Gamma =0$, $\protect\mu =1.254$, the solid line); in
the presence of the attractive DD interaction ($\Gamma =-0.5$, $\protect\mu %
=0.8$, the dashed line); in the presence of the repulsive DD interaction ($%
\Gamma =0.35$, $\protect\mu =1.4$, the dotted line). In all the cases, the
strength of the optical-lattice potential (shown by gray lines) is $V_{0}=1$%
, and the soliton's norm is $P=1.173$. }
\label{fig2}
\end{figure}

Figure \ref{fig2} displays typical examples of fundamental solitons, with
equal values of the norm, which were obtained as numerical solutions to Eq. (%
\ref{stationary}) in cases of zero, attractive, and repulsive DD
interactions. In the presence of the attractive DD interaction, the solitons
are narrower and feature a higher amplitude, while in the case of the DD
repulsion, they are wider, and have a lower amplitude than the respective
soliton in the absence of the DD interaction. The peculiarity of the NPSE in
the case of the attractive contact interaction is that the amplitude of the
soliton is limited by a critical value. The collapse setting in when the
amplitude attains that value \cite{sala1,various}. In contrast to that, in
the ordinary 1D GPE the amplitude may grow indefinitely without causing the
collapse.

If the local interaction is repulsive ($\aleph =-1$), fundamental solitons
in the NPSE may only be supported by the attractive DD interaction, which
must dominate over the contact repulsion. In that case, the difference
between the NPSE and ordinary 1D GPE is not significant. In particular, as
the collapse instability is absent, unstable fundamental solitons can only
evolve into breathers. The situation \ in the case of the repulsion is
similar to that in the case of the discrete NPSE \cite{nash}.

\section{The influence of the dipole-dipole interactions on the collapse}

In this section we focus on the dynamics of fundamental solitons in the
presence of the attractive contact interaction ($\aleph =+1$) and nonlocal
DD of either sign. The collapse occurs in this case, as well as in the
respective discrete setting \cite{nash}.

\begin{figure}[tbp]
\center\includegraphics [width=7.7cm]{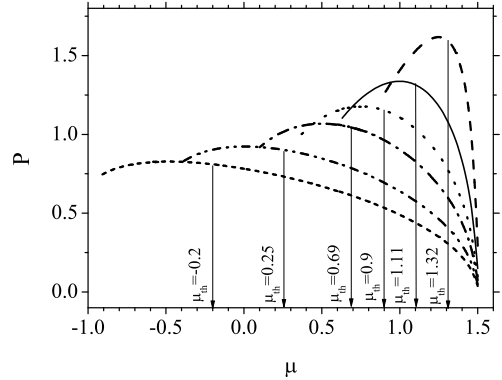}
\caption{$P(\protect\mu )$ diagrams for fundamental solitons in the case of
attractive contact and repulsive, zero, or attractive DD interactions: $%
\Gamma =0.5$ (the dashed line), $\Gamma =0$ (the solid line), $\Gamma =-0.5$
(the dotted line), $\Gamma =-1$ (the dash-dotted line), $\Gamma =-2$ (the
dashed-dotted-dotted line), $\Gamma =-3$ (the short-dash line). In all the
cases, $V_{0}=1$. Values of $\protect\mu _{\mathrm{th}}$ shown in the plots
designate thresholds of the collapse instability.}
\label{fig3}
\end{figure}

A global characteristic of soliton families is the $P(\mu )$ dependence, the
norm versus the chemical potential. In Fig. \ref{fig3} we display the $P(\mu
)$ curves obtained for different values of $\Gamma $. The stability of the
soliton may be assessed according to two different conditions: $dP/d\mu \leq
0$ [the \emph{Vakhitov-Kolokolov} (VK) criterion], and the absence of
eigenvalues for small perturbations with positive real parts (the spectral
condition), see Ref. \cite{stability} and references therein.

In the presence of the attractive contact interaction, the $P(\mu )$
dependences feature two different regions, one where the VK criterion is
fulfilled, and another one, where it is violated. The region where the VK
criterion holds expands/shrinks with the increase of the strength of the
attractive/repulsive DD interactions. These results are in qualitative
agreement with findings for on-site solitons in the discrete version of 1D
NPSE \cite{nash}.

\begin{figure}[tbp]
\center\includegraphics [width=7.7cm]{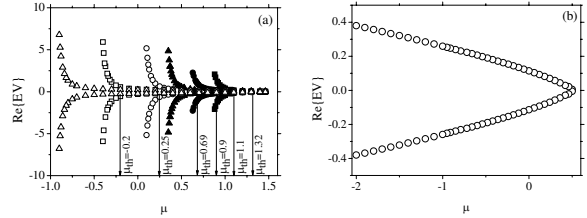}
\caption{The real parts of eigenvalues for small perturbations around the
fundamental soliton versus $\protect\mu $, for the NPSE (a) and cubic GPE
(b). The contact and DD interactions are both attractive in this case.
Parameters are $V_{0}=1$, and (a) $\Gamma =0.5$ (squares), $\Gamma =0$
(circles), $\Gamma =-0.5$ (triangles), $\Gamma =-1$ (empty circles), $\Gamma
=-2$ (empty squares), $\Gamma =-3$ (empty triangles), or (b) $\Gamma =-0.5$.
Values of $\protect\mu _{\mathrm{th}}$ in (a) designate the threshold of the
collapse instability.}
\label{fig4}
\end{figure}

\begin{figure}[tbp]
\center\includegraphics [width=7.7cm]{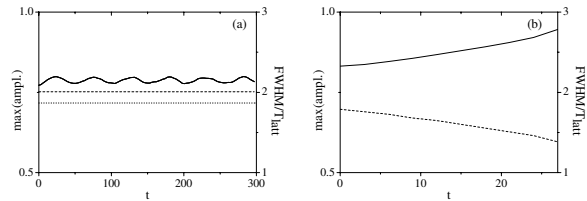}
\caption{Illustration of the evolution of unstable fundamental solitons into
breathers, or onset of the collapse. In the former case (a) the oscillating
amplitude of the breather, and in the latter case (b) the monotonously
growing soliton amplitude are plotted by the solid curves. Dashed and dotted
lines in (a) depict the smallest and largest values of the FWHM (full width
at half maximum) for the breather , while in (b) the FWHM of the collapsing
soliton is depicted by the dashed line. In all the cases the unit of FWHM is
the period of the optical lattice period, $T_{\mathrm{latt}}$ and $V_0=1$. }
\label{fig5}
\end{figure}

The spectral stability condition was examined by numerically computing the
corresponding eigenvalues, using linearized equations for small
perturbations. It has been found that the spectral condition is \emph{%
violated} in the \emph{entire parameter space} which was explored, see Fig. %
\ref{fig4}. The eigenvalue spectra are quite similar to those found for
inter-site solitons in the discrete NPSE with the DD interaction, at large
values of the inter-site coupling constant $C$, which corresponds to the
quasi-continuum limit \cite{nash}. Note that the spectra for solitons in the
NPSE exhibit an abrupt (quasi-exponential) growth of real parts of the
eigenvalues [see Fig. \ref{fig4}(a)], which does not happen in the
respective GPE with the cubic nonlinearity, cf. Fig. \ref{fig4}(b). An
interesting fact revealed by the numerical analysis is that the threshold of
the abrupt growth of the real part of the complex eigenvalues corresponds to
the onset of the collapse instability of the solitons. Indeed, direct
simulations of Eq. (\ref{NPSE1}) show that unstable solitons with $\mu $
higher than the threshold value evolve into robust localized breathers [Fig. %
\ref{fig5}(a)], while the solitons with $\mu $ taken below the threshold
exhibit the collapse instability, which manifests itself through
simultaneous decrease of the soliton's width and growth of the amplitude
towards the limit value, as seen in Fig. \ref{fig5}(b).\ Points of the onset
of the collapse instability are also marked in Fig. \ref{fig3}. It is worthy
to note that the instability sets in earlier than the norm attains the
maximum values (full curves $P(\mu )$ could be drawn in spite of the
instability, as they were obtained from the numerical solution\ of
stationary equation (\ref{stationary})).

\begin{figure}[tbp]
\center\includegraphics
[width=7.7cm]{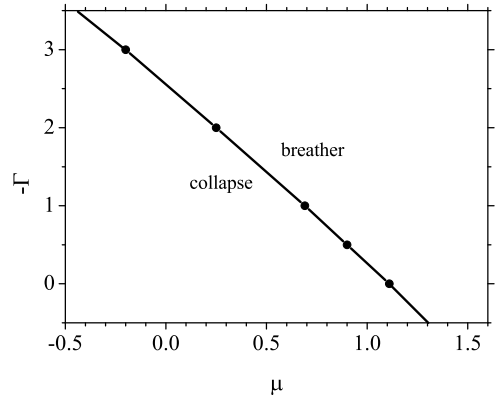}
\caption{The collapse diagram for fundamental solitons in the NPSE with the
attractive sign of both the contact and DD interactions ($V_0=1$). The plotted line
separates parameter regions where solitons evolve into breathers, and
regions where solitons suffer the collapse.}
\label{fig6}
\end{figure}

With the increase of the strength of the attractive DD interaction, the
threshold value of $\mu $ becomes lower, hence the region in the parameter
space where unstable solitons do not collapse but rather evolve into
breathers expands. These results are presented as the \textit{collapse
diagram} in the $(\Gamma ,\mu )$ plane (see Fig. \ref{fig6}), where the
border line separates the collapse region from that where the formation of
breathers takes place. Qualitatively, the nearly linear dependence of the
threshold on $\Gamma (\mu)$ can be understood, assuming that the nonlocality
range is very large, covering the entire soliton. Indeed, in that case the
last term in Eq. (\ref{stationary}) is proportional to $\Gamma P u(z)$,
which implies a linear shift of $\mu $.

Thus, the collapse instability of the fundamental solitons may be suppressed
by using sufficiently strong attractive DD interactions. This prediction of
the stability analysis based on the computation of the eigenvalues was fully
corroborated by direct simulations of Eq. (\ref{NPSE1}).

\section{Conclusion}

The purpose of this work was to achieve a better understanding on the
influence of the nonlocal DD (dipole-dipole) interactions on the stability
and collapse of localized nonlinear modes in the BEC trapped in a
combination of a tight transverse parabolic potential and a relatively loose
periodic OL potential acting in the axial direction. To this end, we have
introduced the model based on the one-dimensional NPSE (nonpolynomial Schr%
\"{o}dinger equation), which includes the contact and DD interaction terms,
as well as the OL potential. Both attractive and repulsive signs of the
contact and DD interactions were considered. The analysis was focused on
fundamental solitons in the semi-infinite gap. While all the stationary
solitons are unstable, an essential conclusion is that the \emph{attractive}
DD interactions may prevent the collapse instability of the fundamental
solitons, replacing the onset of the collapse by the transformation of the
solitons into robust breathers. This general result is consistent with
findings recently reported for the 1D discrete version of the NPSE.

\section*{Acknowledgments}

G.G., A.M. and Lj.H. acknowledge support from the Ministry of Science,
Serbia (through project 141034). The work of B.A.M. was supported, in a
part, by the German-Israel Foundation, through grant No. 149/2006. This
author appreciates hospitality of the Vin\v{c}a Institute of Nuclear
Sciences (Belgrade, Serbia).

\section*{Figures and figure captions}

\end{document}